\documentclass{article}
\usepackage{hiph-preprint}
\usepackage{graphicx}
\volnumber{22} \issuenumber{1} \edyear{2005}                             
\frompage{000} \topage{000}                                              
\recrevdate{1 November 2005}                                              
\def\rightmark{Strange Particle Production in $p+p$ at RHIC}
\def\leftmark{R.Bellwied}

\title{Strange Particle Production Mechanisms in Proton-Proton Collisions at
RHIC}

\authors{
{R.Bellwied$^1$ for the STAR collaboration$^{2}$ %
\index{Bellwied, R.}
}\\[2.812mm]
{\normalsize \hspace*{-8pt}$^1$ Physics Department, Wayne State
University, 666 West Hancock,\\ Detroit, MI 48201, U.S.A., e-mail:
bellwied@physics.wayne.edu\\[0.2ex]
\hspace*{-8pt}$^2$ For the full list of STAR authors and
acknowledgements see appendix 'Collaborations' of this volume
}}

\abstract{We present data on strange particle production in
elementary proton-proton collisions at RHIC energies. Comparison to
leading order and next-to-leading order (NLO) calculations shows
that the fragmentation process is flavor dependent and that higher
order corrections are needed to describe all spectra, in particular
at these collision energies, which are modest compared to those at
the Tevatron. A model (EPOS) which takes into account multiple
scattering between projectile constituents seems to describe the
data best.}

\keyword{proton-proton collisions, higher order calculations}

\PACS{13.87.Fh,12.38.Bx}

\makeindex
\begin{document}

\maketitle

\section{Introduction}
Particle production in elementary proton-proton collisions is widely
recognized as a well understood problem in the string fragmentation
process. The universality of the fragmentation process between
e$^{+}$e$^{-}$ and $p+p$ collisions has been recently confirmed by
Kniehl, Kraemer and Poetter (KKP) in a detailed paper based on pion
and charged hadron production \cite{kkp}. Using the factorization
theorem with the proper next-to-leading order corrections, the
calculation seems to describe the pion and charged hadron production
in RHIC $p+p$ data well \cite{star1,star2}. We therefore expect the
strange particle production to be also described by the standard
fragmentation descriptions. The measurements presented here are
based on two STAR Ph.D.theses \cite{heinz,adams}.

\section{Comparison to PYTHIA} The most commonly applied model available for
the description of hadron-hadron collisions is PYTHIA.  In Fig.1 we
have used PYTHIA (v6.221 with default settings, MSEL=1) \cite{py62}
for comparison to strange particle p$_{T}$ spectra measured in $p+p$
collisions at $\sqrt{s}$=200 GeV (solid line). The model describes
neither the strange meson nor the baryon production. A more recent
version of PYTHIA (v6.317), which includes an improved description
of partonic multiple scattering processes, does well on the strange
meson production (dashed line). The agreement of the model with the
strange baryon spectra improved only when the K-factor was raised to
three (dotted line).

\begin{figure}[hbtl]
\begin{center}
\includegraphics[width=5.in]{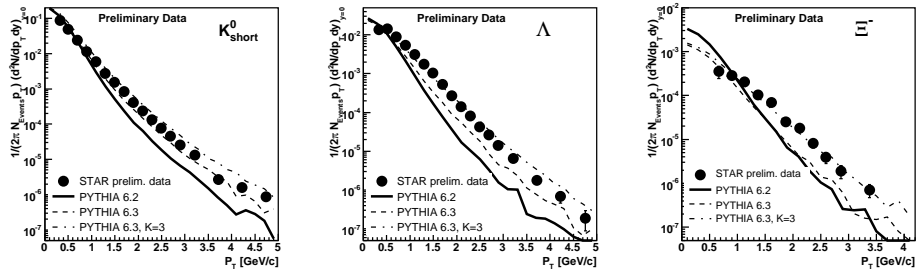}
\caption{K$^{0}_{S}$, $\Lambda$ and $\Xi$ p$_{T}$ spectra compared
to PYTHIA calculations.} \label{fig:pythia}
\end{center}
\end{figure}

\begin{figure}[hbtl]
\begin{center}
\includegraphics[width=5.in]{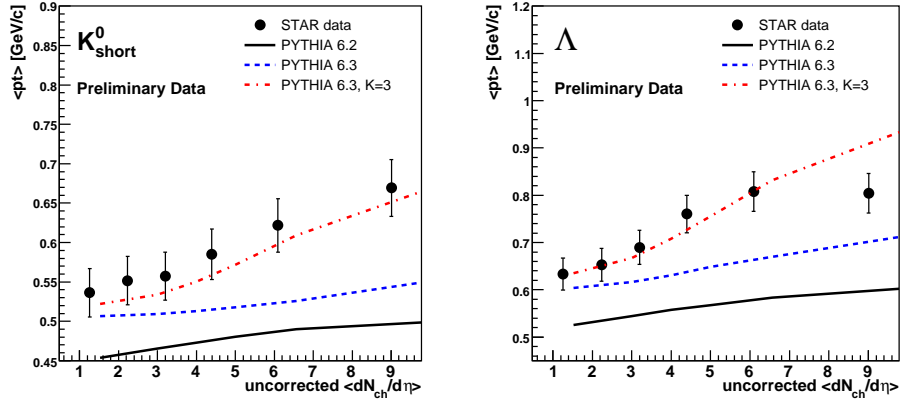}
\caption{K$^{0}_{S}$ and $\Lambda$ $<$p$_{T}$$>$ as a function of
measured charged particle multiplicity per unit pseudo-rapidity
compared to PYTHIA calculations.} \label{fig:pythia-pt}
\end{center}
\end{figure}

The multiplicity dependence of the $<$p$_{T}$$>$ also could not be
reproduced using the default setting, but only when the K-factor was
raised (Fig.2).

Is this large value for K unphysical? Does it signal new physics in
strangeness production? The K-factor, which quantifies the
importance of next-to-leading order (NLO) effects, should increase
from higher to lower collision energies, because the particle
production is less and less dominated by the initial hard two-body
scattering. Eskola et al. showed a K-factor excitation function in
QM02 \cite{eskola}. His calculation is in good agreement with the
large K-factor necessary to describe our data. Therefore, one would
expect actual NLO calculations to perform better than the default
PYTHIA for strange particle production.

\section{Comparison to NLO calculations}

Werner Vogelsang used the KKP parametrization \cite{kkp}, which was
successfully applied to the PHENIX $\pi^{0}$ and the STAR charged
hadron spectra, plus a specific $\Lambda$ fragmentation function by
Vogelsang and DeFlorian \cite{deflorian} in order to describe the
strange particle data shown here. Fig.3 shows that these NLO
calculations describe neither the shape nor the magnitude of the
strange particle production very well, and that the disagreement is
worse for the baryons. The systematic uncertainty of the calculation
is taken into account through varying the renormalization factor,
$\mu$.

\begin{figure}[hbtl]
\begin{center}
\includegraphics[width=5.in]{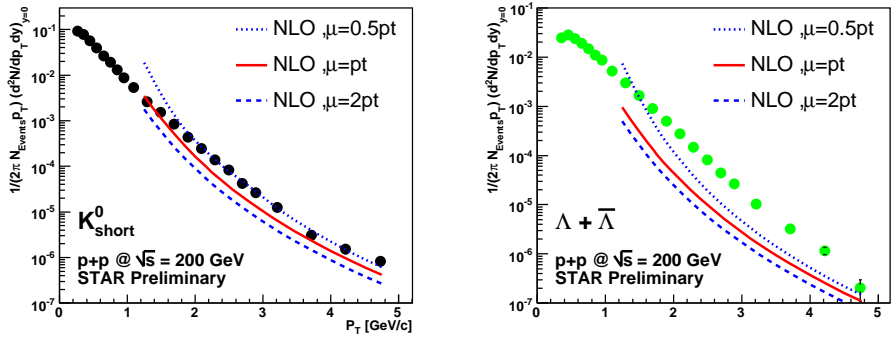}
\caption{K$^{0}_{S}$ and $\Lambda$ p$_{T}$ spectra compared to NLO
calculations based on standard fragmentation functions
\cite{kkp,deflorian}.} \label{fig:nlo}
\end{center}
\end{figure}

The agreement gets significantly better when quark-separated
contributions to the fragmentation function are used according to
Albino, Kniehl and Kraemer \cite{akk} and Bourelly and Soffer
\cite{bs}. It is interesting to note that apparently the quark
separation is more important at the lower (RHIC) energies than at
the higher (CERN) energies \cite{akk}. In addition, the heavy quark
contribution to the light baryon production is not negligible
\cite{bs}, which can be interpreted as a higher order extension to
the quark-separated contributions of the valence quarks.

\section{Comparison to EPOS} Finally we compare to the new EPOS model
\cite{werner}, which takes into account soft and hard partonic
interactions between the produced partons (inner contribution) and
partonic remnants of projectile and target, i.e. diquarks in the
case of proton-proton interactions (outer contributions). The
agreement with the data is remarkable (Fig.4) and is mostly
attributed to soft 'inner' parton cascading, which is a description
of the multiple scattering process of the off-shell partons produced
in the initial parton-parton collision. Surprisingly this soft
contribution is more relevant to the heavy strange baryons than, for
example, the pions \cite{werner}.

\begin{figure}[hbtl]
\begin{center}
\includegraphics[width=5.in]{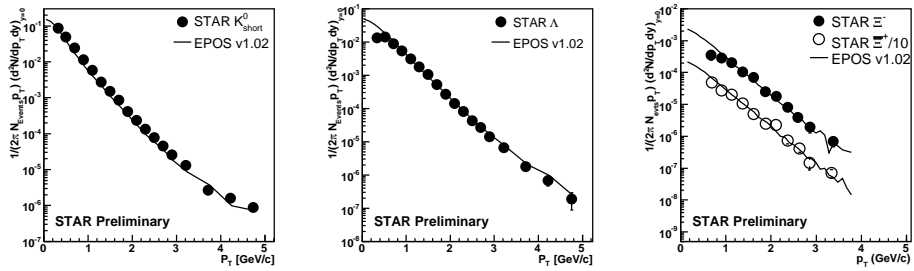}
\caption{K$^{0}_{S}$, $\Lambda$ and $\Xi$ p$_{T}$ spectra compared
to EPOS calculations.} \label{fig:epos}
\end{center}
\end{figure}

\vspace{-0.5cm}
\section{Summary} The good agreement between data and the most recent
extended NLO calculations, as well as the EPOS calculations, points
at the necessity of higher order corrections to the simple string
fragmentation picture for the production of strange baryons and
mesons in $p+p$ collisions. The contributions of each quark flavor
to the fragmentation has to be taken into account separately, and
effects from partonic projectile remnants or cascading soft
non-valence quarks are non-negligible. Therefore, strange particles,
which are produced abundantly at RHIC in $p+p$ collisions, provide a
good test for the quantitative determination of the generation
mechanism in elementary collisions, where apparently simple string
fragmentation is not sufficient to describe all of the hadron
production.

\end{document}